\documentclass[journal]{IEEEtran}
\usepackage{algorithm}
\usepackage{algpseudocode}
\usepackage{amsmath}
\usepackage{newtxmath}
\usepackage{amsfonts}
\usepackage{multirow}
\usepackage{balance}
\usepackage{placeins}
\usepackage[colorlinks=true, allcolors=black]{hyperref}
\ifCLASSINFOpdf
   \usepackage[pdftex]{graphicx}
\else
\fi
  \usepackage[caption=false,font=footnotesize]{subfig}
\begin{document}
%
% paper title
% Titles are generally capitalized except for words such as a, an, and, as,
% at, but, by, for, in, nor, of, on, or, the, to and up, which are usually
% not capitalized unless they are the first or last word of the title.
% Linebreaks \\ can be used within to get better formatting as desired.
% Do not put math or special symbols in the title.
\title{CF-Net: A Cross-Feature Reconstruction Network for High-Accuracy 1-Bit Target Classification}
%
%
% author names and IEEE memberships
% note positions of commas and nonbreaking spaces ( ~ ) LaTeX will not break
% a structure at a ~ so this keeps an author's name from being broken across
% two lines.
% use \thanks{} to gain access to the first footnote area
% a separate \thanks must be used for each paragraph as LaTeX2e's \thanks
% was not built to handle multiple paragraphs
%

\author{Jundong~Qi,
        Weize~Sun\thanks{Corresponding author: Weize Sun (proton198601@hotmail.com)},
        Shaowu~Chen,
        Lei~Huang,
        and~Qiuchen~Liu% <-this % stops a space
\thanks{The authors are with the College of Electronics and Information Engineering, Shenzhen University, Shenzhen, China.}}

\maketitle

\begin{abstract}
Target classification is a fundamental task in radar systems, and its performance critically depends on the quantization precision of the signal. While high-precision quantization (e.g. 16-bit) is well established, 1-bit quantization offers distinct advantages by enabling direct sampling at high frequencies and eliminating complex intermediate stages. However, its extreme quantization leads to significant information loss.
Although higher sampling rates can compensate for this loss, such oversampling is impractical at the high frequencies targeted for direct sampling. 
To achieve high-accuracy classification directly from 1-bit radar data \textbf{under the same sampling rate}, this paper proposes a novel two-stage deep learning framework, CF-Net. 
First, we introduce a self-supervised pre-training strategy based on a dual-branch U-Net architecture. This network learns to restore high-fidelity 16-bit images from their 1-bit counterparts via a cross-feature reconstruction task, forcing the 1-bit encoder to learn robust features despite extreme quantization. Subsequently, this pre-trained encoder is repurposed and fine-tuned for the downstream multi-class target classification task.
Experiments on two radar target datasets demonstrate that CF-Net can effectively extract discriminative features from 1-bit imagery, achieving comparable and even superior accuracy to some 16-bit methods without oversampling.
 % paving the way for efficient, real-time radar applications.
\end{abstract}

% Note that keywords are not normally used for peerreview papers.

% For peer review papers, you can put extra information on the cover
% page as needed:
% \ifCLASSOPTIONpeerreview
% \begin{center} \bfseries EDICS Category: 3-BBND \end{center}
% \fi
%
% For peerreview papers, this IEEEtran command inserts a page break and
% creates the second title. It will be ignored for other modes.
\IEEEpeerreviewmaketitle

%%% ========== START: SECTION I INTRODUCTION (FINAL, CONFIRMED CORRECT) ========== %%%
\section{Introduction}
\label{sec:introduction}

\IEEEPARstart{R}{adar} sensors are widely used in maritime surveillance, autonomous navigation, environmental monitoring, and human–object interaction due to their all-weather reliability and robustness to changes in illumination \cite{hou2020fusar, jin2022uwb}. With the growing adoption of millimeter-wave radar on resource-constrained platforms such as drones, vehicles, and low-power embedded devices, the demand for compact and energy-efficient sensing pipelines has increased substantially \cite{chen2016target, zhao2019one}. These constraints highlight the importance of reducing the volume of data while maintaining the interpretability required for high-level perception tasks.

Traditional radar systems typically operate with high-precision quantization (e.g., 16-bit), generating large amounts of data and imposing notable burdens on storage, transmission, and computation \cite{chen2016target}. To address this issue, 1-bit quantization has attracted considerable interest for its ability to compress radar echoes, simplify ADC circuitry, and lower system power consumption \cite{zhao2019one, xiao2022one, si2023convolutional, guo2024residual}. It has also been explored in high-frequency sampling scenarios where conventional ADCs become difficult to deploy \cite{wu2024one}. However, extreme reduction in bit depth inevitably leads to substantial information loss, including the approximate \(2\,\mathrm{dB}\) degradation of the signal-to-noise ratio reported in \cite{xiao2022one}, which poses major challenges for downstream interpretation.

The impact of this information loss becomes particularly pronounced in high-level semantic tasks such as target classification. Features extracted directly from 1-bit measurements are often unstable and are easily corrupted by quantization artifacts, making deep networks prone to learning spurious patterns instead of meaningful target characteristics. Existing studies on 1-bit radar processing mainly focus on low-level recovery or imaging enhancement—such as adaptive thresholds, sparsity-driven models, and improved reconstruction algorithms \cite{zhao2019one, wu2024one}—which are useful for visualization but do not resolve the difficulty of learning discriminative representations suitable for classification. Consequently, classification under the same sampling rate and antenna configuration as full-precision systems remains an open problem.

This motivates a different perspective: instead of treating 1-bit radar as an isolated data representation, we aim to leverage the rich semantic cues available in high-precision 16-bit data and transfer them into the 1-bit feature space. High-bit-depth radar images inherently contain more stable and informative structures, and aligning 1-bit features with these structures has the potential to compensate for quantization-induced degradation without modifying the radar hardware or sampling pipeline. This idea is closely related to recent advances in cross-feature learning, representation alignment, and knowledge transfer, which have demonstrated strong effectiveness in remote sensing applications \cite{wang2023cross, shi2024unsupervised, zheng2023partial}.
To leverage this potential, we propose our {\textbf{Cross-Feature Network (CF-Net)}}.

The main contributions of this paper are summarized as follows:
\begin{itemize}
    \item We propose a novel two-stage training paradigm (CF-Net). The core idea is to design a self-supervised cross-feature reconstruction pretext task, which uses information-rich 16-bit data as supervision to force an encoder to learn robust and semantically-aligned feature representations from extremely quantized 1-bit data.
    
    \item We design the specific network architecture to implement this paradigm. During pre-training, a dual-branch U-Net structure is used, incorporating a Cross-Attention module to facilitate feature interaction and alignment between the 1-bit (student) and 16-bit (teacher) branches. During fine-tuning, a multi-scale feature fusion strategy is employed to capture discriminative information at different levels.
    
    \item We introduce a compound loss function to guide the pre-training. In addition to the primary reconstruction ($L_{\mathrm{rec}}$) and consistency ($L_{\mathrm{con}}$) losses, it innovatively incorporates a Feature Alignment Loss ($L_{\mathrm{align}}$) and a Feature Separation Loss ($L_{\mathrm{sep}}$). This design ensures the learned feature space is not only high-fidelity for reconstruction but also highly discriminative for classification.
\end{itemize}

%%% ========== START: REVISED REMAINDER OF THE PAPER PARAGRAPH ========== %%%
The remainder of this paper is organized as follows. Section~\ref{sec:related_work} reviews the related work. Section~\ref{sec:methodology} details the architecture and training methodology of our proposed CF-Net. Section~\ref{sec:experiments} presents the experimental results, including the performance on SAR image classification and the generalization of our framework to the 1-bit human activity recognition task. Finally, Section~\ref{sec:conclusion} concludes the paper and discusses future work.

%%%%%%%%%%%%%%%%%%%%%%%%%%%%%%%%%%%%%%%%%%%%%%%%%%%%%%%%%%%%%%%%%%%%%%%%%%%%%%%%
%                                SECTION II
%                               RELATED WORK
%%%%%%%%%%%%%%%%%%%%%%%%%%%%%%%%%%%%%%%%%%%%%%%%%%%%%%%%%%%%%%%%%%%%%%%%%%%%%%%%
\section{Related Work}
\label{sec:related_work}

Deep learning has emerged as a powerful paradigm for intelligent radar information interpretation, enabling automatic feature extraction and demonstrating superior performance across various tasks beyond traditional SAR applications \cite{zhao2023cubelearn}. Radar systems, operating across different frequency bands like millimeter-wave or Ultra-Wideband (UWB), are increasingly employed for fine-grained perception. Common objectives include human activity recognition \cite{jin2022uwb, sainath2015convolutional, hazra2019short}, fall detection \cite{yao2022fall}, and complex video understanding using advanced spectrum-spatial-temporal attention mechanisms \cite{zhang2024sat}. Recent trends even explore generative artificial intelligence for diverse interpretation scenarios \cite{zhu2025generative}. However, deploying these advanced models on resource-constrained platforms (e.g., drones, edge devices) imposes stringent efficiency requirements, necessitating specialized system-on-chip (SoC) architectures and highly efficient algorithms \cite{xie2023advancements}.

In the specific domain of SAR target classification, deep learning has achieved state-of-the-art results, predominantly on high-precision (e.g., 16-bit) data \cite{hou2020fusar, chen2016target}. To further boost performance, research has shifted towards more sophisticated architectures. Complex-valued networks have been introduced to better leverage phase information \cite{ren2023complex}. Various Transformer-based models, including Swin Transformers \cite{zhang2024swin}, multimodal fusion Transformers \cite{roy2023multimodal}, and hierarchical cross-scale Transformers \cite{zhang2025multiple}, have been widely adopted to capture long-range global dependencies. Furthermore, advanced fusion frameworks—such as interactive attention for heterogeneous tensor decomposition \cite{zhou2024interactive}, frequency-domain fusion networks (SFFNet) \cite{zhu2024sffnet}, and lightweight CNN-Transformer hybrids \cite{yan2024cd}—demonstrate the power of integrating multi-source information. Recently, federated learning has also been explored for multi-label classification in remote sensing, addressing data privacy concerns \cite{buyuktas2024transformer}. In addition, improving the robustness against adversarial attacks and backdoors \cite{wei2024moar, zeng2024dacobd} and enabling transferable classification across different satellite conditions \cite{zhao2022transferable} are crucial directions in SAR image classification.

Despite these successes in the high-precision domain, applying deep learning to 1-bit radar data presents unique challenges due to severe information loss. Existing research in the 1-bit domain has focused heavily on low-level signal processing to mitigate quantization artifacts during imaging. Foundational works analyzed performance degradation and established basic imaging principles \cite{xiao2022one, zhao2019one, wu2024one}. Subsequent research proposed various optimization techniques to improve image quality, such as using time-varying thresholds \cite{demir2018one}, sparse logistic regression \cite{ge2023sparse}, and total variation regularization \cite{niu2023one}. More recently, advanced methods leveraging frequency agility \cite{li2023maximizing}, structured sparsity to mitigate sign flips \cite{zhang2024enhanced}, and slow-time fluctuating thresholds \cite{nie2025one} have further pushed the boundaries of 1-bit reconstruction quality.

Critically, the application of deep learning directly to 1-bit radar data remains nascent. Existing deep learning attempts predominantly focus on reconstruction or restoration rather than high-level understanding. For instance, convolutional networks \cite{si2023convolutional} and attention-augmented U-Nets \cite{guo2024residual} have been used to suppress harmonics and restore 1-bit images. While valuable for visualization, these methods do not directly tackle the challenge of high-accuracy classification under the strict constraint of using the same sampling rate as high-precision systems.

To bridge this gap, our approach draws inspiration from advanced representation learning paradigms. Techniques like multi-scale masked autoencoders (Scale-MAE) \cite{reed2023scale} and super-resolution Transformers \cite{ma2024mswagan} highlight the importance of robust feature hierarchies in challenging remote sensing tasks. We specifically leverage concepts from Knowledge Distillation (KD) \cite{ji2022knowledge}, which has evolved from simple model compression to sophisticated tasks like logit standardization \cite{sun2024logit} and task-specific distillation from large models \cite{jang2024vl2lite}. Notably, RadarDistill \cite{bang2024radardistill} successfully applied cross-feature KD to boost radar object detection using LiDAR features, validating this direction. Furthermore, ideas from Siamese networks and consistency learning, often used for object tracking \cite{bertinetto2016fully} or handling noisy data in medical imaging \cite{qiu2024noise, qiu2025noise}, and recent efficient pre-training methods using Siamese cropped masked autoencoders \cite{eymael2024efficient} and joint alignment and regression for video grounding \cite{tan2024siamese}, provide a solid foundation for learning invariant representations. Inspired by these, we propose a cross-feature reconstruction framework that uses high-fidelity data to guide the learning of a robust 1-bit encoder, enabling high-accuracy classification despite extreme quantization.

%%%%%%%%%%%%%%%%%%%%%%%%%%%%%%%%%%%%%%%%%%%%%%%%%%%%%%%%%%%%%%%%%%%%%%%%%%%%%%%%
%                                SECTION III
%                       THE PROPOSED CMR-NET FRAMEWORK
%%%%%%%%%%%%%%%%%%%%%%%%%%%%%%%%%%%%%%%%%%%%%%%%%%%%%%%%%%%%%%%%%%%%%%%%%%%%%%%%
\section{The Proposed CF-Net}
\label{sec:methodology}

In this section, we present the proposed CF-Net. We first provide an overview of our two-stage framework, which is conceptually grounded in cross-feature knowledge distillation. Then, we detail the network architecture and the compound loss function for the pre-training stage. Finally, the fine-tuning process for classification is described.

% --- 你的架构图代码放在这里 ---
\begin{figure*}[t!]
\centering
% 注意：这里用 figure* 和 \textwidth 是为了让图跨双栏显示，更清晰。如果想单栏显示，用 figure 和 \columnwidth
\includegraphics[width=0.7\textwidth]{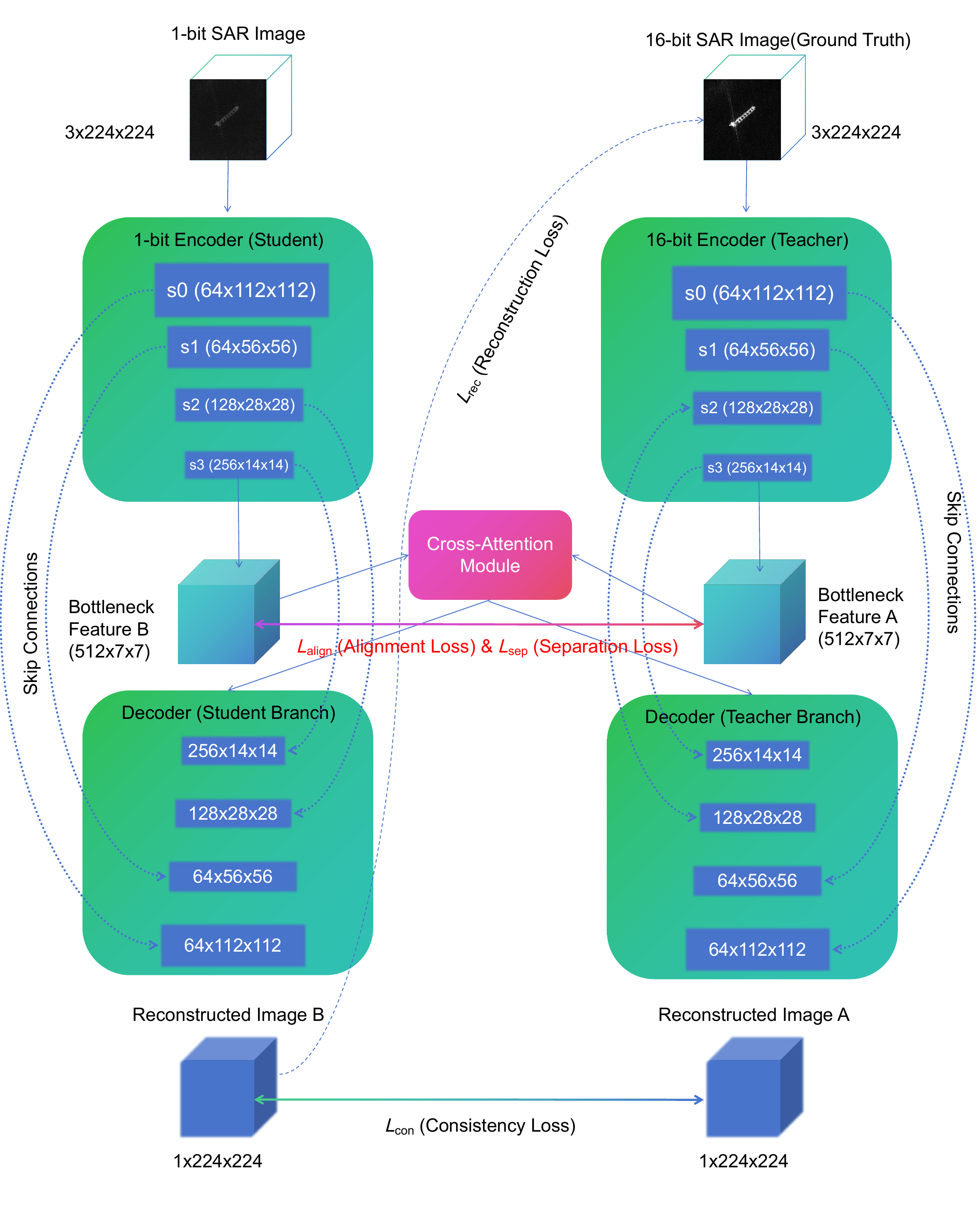} 
\caption{The pre-training Stage-1 architecture of the proposed CF-Net. It features a symmetrical dual-branch U-Net structure: the student branch learns to reconstruct high-fidelity images from 1-bit data under the guidance of a teacher branch, interacting via a central cross-attention module. The four key loss functions that guide the training are also shown.}
\label{fig:cf_net_arch}
\end{figure*}
% --- 架构图代码结束 ---
% ==================== 请用下面的代码块替换你原来的 Framework Overview 子章节 ====================
\subsection{Framework Overview}
The core challenge in 1-bit data classification is to extract discriminative features from an information-impoverished data source. To overcome this, our CF-Net framework adopts a two-stage learning strategy based on the principle of knowledge distillation \cite{ji2022knowledge}.

    \textbf{Stage 1: Self-Supervised Cross-Feature Pre-training.} 
    In this stage, a dual-branch U-Net architecture acts as our knowledge distillation engine. 
    The 16-bit data stream serves as the ``teacher'', providing rich, clean feature targets. 
    The 1-bit data stream acts as the ``student''. 
    The network is trained on a pretext task of reconstructing the 16-bit image from the 1-bit image. 
    A carefully designed compound loss function forces the ``student'' encoder to not just reconstruct pixels, but to mimic the feature space of the ``teacher'', thereby learning a robust and semantically rich representation.
    
    \textbf{Stage 2: Supervised Classification Fine-tuning.} 
    After pre-training, the distilled knowledge is encapsulated in the 1-bit encoder. 
    We detach this ``student'' encoder and repurpose it as the backbone for a classification network. 
    A classification head is appended, and the model is fine-tuned on the 1-bit information dataset with class labels.

The detailed architecture for this crucial pre-training stage is illustrated in Fig.~\ref{fig:cf_net_arch}.

\subsection{Stage 1: Cross-Feature Reconstruction Pre-training}
This stage is the cornerstone of our framework, designed to enable the 1-bit encoder to deeply understand radar target structures.

\subsubsection{Network Architecture}
The core of the pre-training stage is a dual-branch U-Net, architected to facilitate knowledge transfer from high-fidelity 16-bit data (teacher) to 1-bit data (student). Key components include:

\begin{itemize}
    \item \textbf{Dual Encoders:} 
    The network uses two identical and independent encoders, $E_{1\text{bit}}$ (student branch) and $E_{16\text{bit}}$ (teacher branch). Each encoder is built upon a ResNet-34 backbone pre-trained on ImageNet to extract hierarchical features. This backbone provides a strong foundation for feature extraction due to its proven performance.
    
    \item \textbf{Cross-Attention Fusion:} 
    At the U-Net bottleneck (where features are most concentrated), we introduce a Cross-Attention module to enable interaction between the two branches. This module follows the principles of the transformer architecture \cite{vaswani2017attention}, utilizing Query (Q), Key (K), and Value (V) projections. Specifically, the feature map from the student encoder ($f_{1\text{bit}}$) acts as the Query, while the features from the teacher encoder ($f_{16\text{bit}}$) provide the Key and Value. This allows the student branch to selectively ``attend to'' the teacher’s feature representations and query the most salient information from the teacher branch, forcing it to focus on reconstructing essential structures and suppressing noise, which is a critical mechanism for effective knowledge distillation.
    
    \item \textbf{Dual Decoders:} Symmetrical to the encoders, two parallel decoders, $D_{1\text{bit}}$ and $D_{16\text{bit}}$, are used to reconstruct the images. They receive fused feature from the cross-attention module and use skip connections from their respective encoders to recover spatial details lost during downsampling.
\end{itemize}

\begin{algorithm}[t!]
\caption{CF-Net Pre-training Stage}
\label{alg:pretraining}
\begin{algorithmic}[1]
\State \textbf{Input:} Training dataset $D_{\mathrm{train}} = \{(x_{\mathrm{1bit}}^{(i)}, x_{\mathrm{16bit}}^{(i)}, y^{(i)})\}_{i=1}^{N}$
\State \textbf{Input:} CF-Net model $\mathcal{F}$ with student branch $\mathcal{F}_{\mathrm{S}}$ and teacher branch $\mathcal{F}_{\mathrm{T}}$
\State \textbf{Input:} Loss weights $\lambda_{\mathrm{rec}}, \lambda_{\mathrm{con}}, \lambda_{\mathrm{align}}, \lambda_{\mathrm{sep}}$
\State Initialize parameters of $\mathcal{F}$
\For{each epoch}
    \For{each batch $(X_{\mathrm{1bit}}, X_{\mathrm{16bit}}, Y)$ in $D_{\mathrm{train}}$}
        \State \Comment{Forward pass through both branches}
        \State $\hat{X}_{\mathrm{T}}, \hat{X}_{\mathrm{S}}, [f_{\mathrm{T}}, f_{\mathrm{S}}] \gets \mathcal{F}(X_{\mathrm{16bit}}, X_{\mathrm{1bit}})$ 
        
        \State \Comment{Calculate compound loss}
        \State $L_{\mathrm{rec}} \gets \mathrm{MSE}(\hat{X}_{\mathrm{S}}, X_{\mathrm{16bit}})$
        \State $L_{\mathrm{con}} \gets \mathrm{MSE}(\hat{X}_{\mathrm{T}}, \hat{X}_{\mathrm{S}})$
        \State $L_{\mathrm{align}} \gets 1 - \mathrm{CosineSimilarity}(f_{\mathrm{T}}, f_{\mathrm{S}})$
        \State $L_{\mathrm{sep}} \gets \mathrm{TripletLoss}(f_{\mathrm{S}}, Y)$
        \State $L_{\mathrm{total}} \gets 
        \lambda_{\mathrm{rec}} L_{\mathrm{rec}} +
        \lambda_{\mathrm{con}} L_{\mathrm{con}} +
        \lambda_{\mathrm{align}} L_{\mathrm{align}} +
        \lambda_{\mathrm{sep}} L_{\mathrm{sep}}$
        
        \State \Comment{Backward pass and optimization}
        \State Backpropagate $L_{\mathrm{total}}$ and update $\mathcal{F}$'s parameters
    \EndFor
\EndFor
\State \textbf{Output:} Pre-trained 1-bit Encoder (Student) $E_{\mathrm{1bit}}$
\end{algorithmic}
\end{algorithm}

\subsubsection{Compound Loss Function}
The loss function is critical for effective knowledge distillation. Our compound loss $L_{\mathrm{pretrain}}$ is a weighted sum of four components:

\begin{equation}
    L_{\mathrm{pretrain}} =
    \lambda_{\mathrm{rec}} L_{\mathrm{rec}} +
    \lambda_{\mathrm{con}} L_{\mathrm{con}} +
    \lambda_{\mathrm{align}} L_{\mathrm{align}} +
    \lambda_{\mathrm{sep}} L_{\mathrm{sep}}
\end{equation}
Let $x_{\mathrm{1bit}}$ and $x_{\mathrm{16bit}}$ denote the paired 1-bit and 16-bit input data or images. The CF-Net, denoted by $\mathcal{F}$, produces two reconstructed outputs, $\hat{x}_{\mathrm{T}}$ from the teacher branch and $\hat{x}_{\mathrm{S}}$ from the student branch, along with their respective bottleneck features, $f_{\mathrm{T}}$ and $f_{\mathrm{S}}$. The four loss components are then defined as follows:

\begin{itemize}
    \item \textbf{Reconstruction Loss ($L_{\mathrm{rec}}$)}: A primary Mean Squared Error (MSE) loss that ensures the student branch accurately reconstructs the ground truth 16-bit image:
    \begin{equation}
        L_{\mathrm{rec}} = \mathbb{E}\left[ \lVert \hat{x}_{\mathrm{S}} - x_{\mathrm{16bit}} \rVert^2_2 \right]
    \end{equation}

    \item \textbf{Consistency Loss ($L_{\mathrm{con}}$)}: An auxiliary MSE loss that enforces similarity between the outputs of the two decoder branches, which regularizes the training process:
    \begin{equation}
        L_{\mathrm{con}} = \mathbb{E}\left[ \lVert \hat{x}_{\mathrm{T}} - \hat{x}_{\mathrm{S}} \rVert^2_2 \right]
    \end{equation}

    \item \textbf{Feature Alignment Loss ($L_{\mathrm{align}}$)}: Encourages semantic similarity between the student ($f_{\mathrm{S}}$) and teacher ($f_{\mathrm{T}}$) bottleneck features, using cosine similarity for feature-level knowledge transfer:
    \begin{equation}
        L_{\mathrm{align}} = 1 - \frac{f_{\mathrm{T}} \cdot f_{\mathrm{S}}}{\lVert f_{\mathrm{T}} \rVert_2 \cdot \lVert f_{\mathrm{S}} \rVert_2}
    \end{equation}

    \item \textbf{Feature Separation Loss ($L_{\mathrm{sep}}$)}:
    Structures the feature space for better classification using a batch-hard triplet loss on the student encoder's features, a formulation that is consistent with supervised contrastive learning principles~\cite{Khosla2020SupCon}. This loss pulls features of the same class closer while pushing features of different classes apart:
    \begin{equation}
        L_{\mathrm{sep}} =
        \sum_{i=1}^{B}
        \max\bigl(0,\,
        d(f_{\mathrm{a}}^i, f_{\mathrm{p}}^i)
        - d(f_{\mathrm{a}}^i, f_{\mathrm{n}}^i)
        + m \bigr)
    \end{equation}
    where $d(\cdot, \cdot)$ is the Euclidean distance, $m$ is a margin, and $(f_{\mathrm{a}}^i, f_{\mathrm{p}}^i, f_{\mathrm{n}}^i)$ represent the anchor, positive, and negative samples within a batch of size $B$.
\end{itemize}

The overall pre-training process is summarized in Algorithm~\ref{alg:pretraining}.

% ===========================================================================
% ====== 最终完善版：Stage 2: Classification Fine-tuning 子章节 ======
% ===========================================================================
\subsection{Stage 2: Classification Fine-tuning}
\begin{figure}[t!]
\centering
\includegraphics[width=\columnwidth]{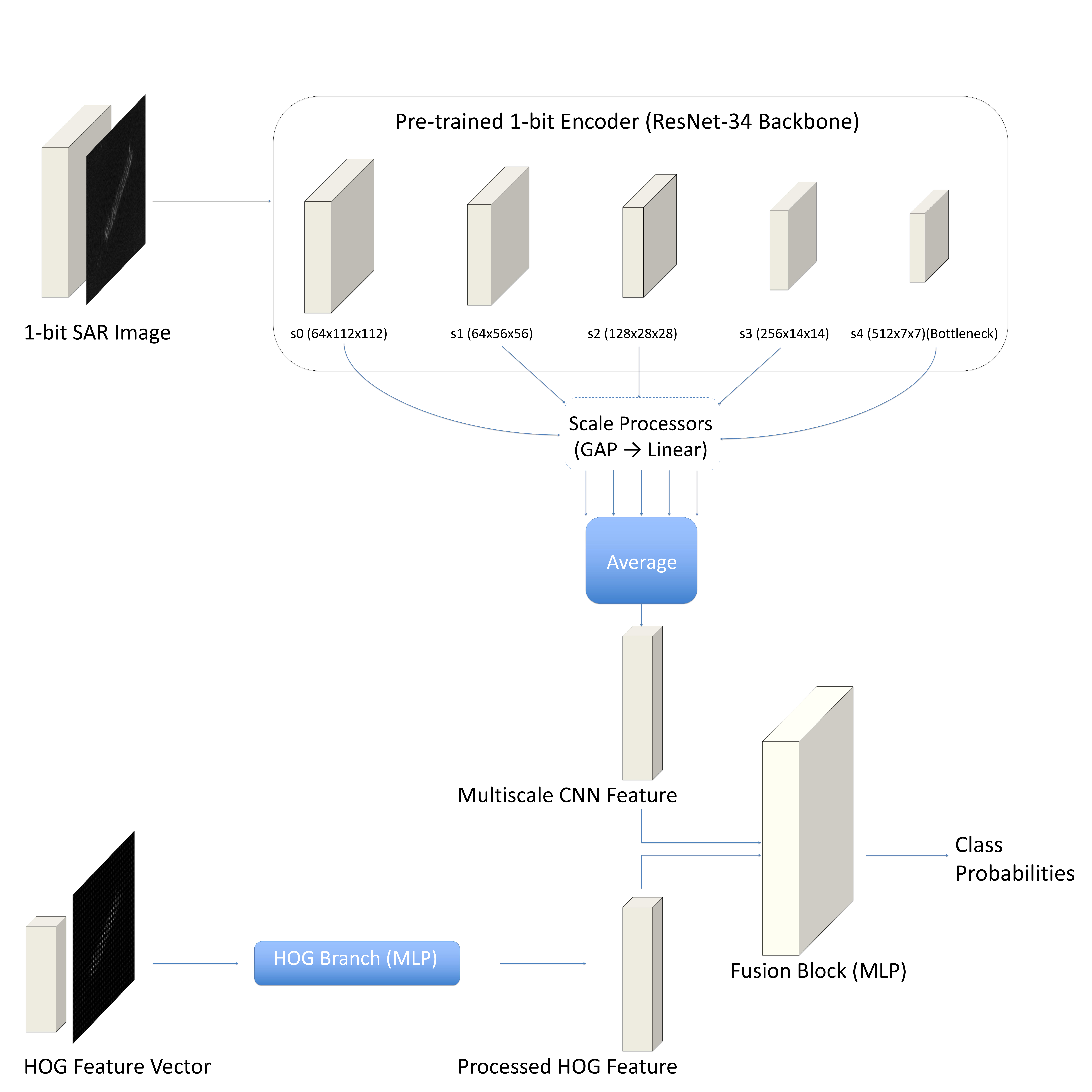}
\caption{Architecture of the proposed multi-scale fusion classification network (Stage 2). The pre-trained encoder extracts features at multiple scales from the 1-bit SAR image. These are aggregated and fused with HOG features before being passed to the final classifier.}
\label{fig:classification_network}
\end{figure}

After pre-training, the student branch's 1-bit encoder ($E_{1\text{bit}}$) has learned a robust feature representation, providing a solid foundation for downstream classification. In many radar-based recognition scenarios, certain human-selected features have proven to be effective in capturing structural patterns that complement deep representations. For example, the histogram of Oriented Gradients (HOG) has shown strong discriminative capability in high-bit-depth SAR classification \cite{zhang2022hog, zhang2021injection}. Inspired by this observation, our classification stage integrates deep multi-scale features learned from the pre-trained encoder with such handcrafted cues, enhancing both robustness and interpretability. The final classification network (illustrated in Fig.~\ref{fig:classification_network}) adopts a dual-branch structure that leverages deep multi-scale representations together with human-selected structural features.

% ==========================================================
% ============ 下游分类网络架构图 (最终版) ====================
% ==========================================================

The network processes two parallel input streams: the 1-bit SAR image and its corresponding HOG feature vector. A critical step (detailed in ablation studies) is that HOG features are extracted not from the noisy 1-bit image, but from the high-fidelity image reconstructed by CF-Net. This ``feature enhancement" step ensures the quality of handcrafted features. The forward pass for the 1-bit SAR image $x_{\mathrm{img}}$ and its corresponding HOG feature vector $x_{\mathrm{hog}}$ is formulated as follows:

1) \textbf{Multi-Scale Feature Extraction:} The pre-trained encoder backbone, $E_{1\text{bit}}$, extracts five feature maps $\{s_0, s_1, s_2, s_3, s_4\}$ from different hierarchical levels of the input image:

% \begin{equation}
%     \{s_0, \dots, s_4\} = E_{1\text{bit}}(x_{img})
% \end{equation}
\begin{equation}
    \{s_0, \dots, s_4\} = E_{\mathrm{1bit}}(x_{\mathrm{img}})
\end{equation}

2) \textbf{Scale Processing and Aggregation:} Each feature map $s_i$ is processed by its dedicated scale processor $P_i$, which consists of global average pooling and a linear layer, to generate a fixed-dimension vector $v_i$. These vectors are then aggregated via element-wise averaging to form a single multi-scale CNN feature vector $v_{\mathrm{cnn}}$:

% \begin{equation}
%     v_{cnn} = \frac{1}{5} \sum_{i=0}^{4} P_i(s_i)
% \end{equation}
\begin{equation}
    v_{\mathrm{cnn}} = \frac{1}{5} \sum_{i=0}^{4} P_i(s_i)
\end{equation}

3) \textbf{HOG Feature Processing:} Concurrently, the HOG feature vector $x_{\mathrm{hog}}$ is processed by a separate MLP (the HOG Branch $H$) to produce a higher-level representation $v_{\mathrm{hog}}$:

% \begin{equation}
%     v_{hog} = H(x_{hog})
% \end{equation}
\begin{equation}
    \mathrm{v}_{\mathrm{hog}} = \mathrm{H}(\mathrm{x}_{\mathrm{hog}})
\end{equation}

4) \textbf{Fusion and Classification:} The multi-scale CNN vector and the processed HOG vector are concatenated (denoted by $\oplus$) and passed through a final fusion block $\mathcal{G}_{\mathrm{fuse}}$ and a classification head $\mathcal{G}_{\mathrm{head}}$ to yield final classification logits:

% \begin{equation}
%     v_{fused} = v_{cnn} \oplus v_{hog}
% \end{equation}
\begin{equation}
    v_{\mathrm{fused}} = v_{\mathrm{cnn}} \oplus v_{\mathrm{hog}}
\end{equation}

% \begin{equation}
%     logits = \mathcal{G}_{head}(\mathcal{G}_{fuse}(v_{fused}))
% \end{equation}
\begin{equation}
    \text{logits} = \mathcal{G}_{\mathrm{head}}\bigl(\mathcal{G}_{\mathrm{fuse}}(v_{\mathrm{fused}})\bigr)
\end{equation}

The multi-scale feature fusion architecture is a critical design choice. 
It is motivated by proven effectiveness in challenging SAR classification scenarios (e.g., few-shot learning), where capturing fine-grained local details and high-level semantic information from limited data is essential for robust, discriminative feature representations \cite{gao2024multibranch}. By aggregating features from different encoder depths, the network better overcomes the information sparsity of 1-bit data.

The entire network is fine-tuned using Focal Loss to handle class imbalance in the dataset. The detailed two-phase fine-tuning strategy is described in Algorithm~\ref{alg:finetuning}.

\begin{algorithm}[t!]
\caption{Fine-tuning for Downstream Classification}
\label{alg:finetuning}
\begin{algorithmic}[1]
\State \textbf{Input:} Training data $(X_{\mathrm{1bit}}, X_{\mathrm{HOG}}, Y)$, Validation data $D_{\mathrm{val}}$
\State \textbf{Input:} Pre-trained 1-bit Encoder $E_{\mathrm{1bit}}$
\State Initialize classification model $\mathcal{G}$ with $E_{\mathrm{1bit}}$ as backbone
\State
\Statex \textit{--- Phase 1: Train classification head only ---}
\State Freeze all layers in $E_{\mathrm{1bit}}$
\For{epoch = 1 to \texttt{FINETUNE\_EPOCHS\_HEAD\_ONLY}}
    \For{each batch $(x_{\mathrm{1bit}}, x_{\mathrm{HOG}}, y)$ in training data}
        \State $p \gets \mathcal{G}(x_{\mathrm{1bit}}, x_{\mathrm{HOG}})$
        \State $L \gets \mathrm{FocalLoss}(p, y)$
        \State Backpropagate and update parameters of $\mathcal{G}$
    \EndFor
\EndFor
\State
\Statex \textit{--- Phase 2: Full network fine-tuning ---}
\State Unfreeze all layers of $\mathcal{G}$
\State Configure optimizer with differential learning rates
\For{epoch = 1 to \texttt{FINETUNE\_EPOCHS\_FULL}}
    \For{each batch $(x_{\mathrm{1bit}}, x_{\mathrm{HOG}}, y)$ in training data}
        \State $p \gets \mathcal{G}(x_{\mathrm{1bit}}, x_{\mathrm{HOG}})$
        \State $L \gets \mathrm{FocalLoss}(p, y)$
        \State Backpropagate and update parameters of $\mathcal{G}$
    \EndFor
    \State Evaluate model on $D_{\mathrm{val}}$ and save the best checkpoint
\EndFor
\State \textbf{Output:} Fine-tuned classification model $\mathcal{G}$
\end{algorithmic}
\end{algorithm}

\subsection{Data Augmentation for Imbalanced Learning}
\label{sec:data_augmentation}
In many radar datasets, the sample distribution is inherently biased due to variations in target occurrence, imaging geometry, and acquisition conditions. Recognizing this imbalance, we design a series of tailored data augmentation strategies to improve model robustness and prevent overfitting. Addressing data scarcity and imbalance is a key research topic in radar target recognition, with approaches ranging from geometric transformations to advanced generative techniques \cite{wang2024textgen}. Following this principle, our training pipeline incorporates class-aware oversampling and diversified augmentations such as random rotations, flips, brightness adjustments, and speckle noise simulation. These operations ensure that minority classes are sufficiently represented during training, while for more balanced datasets, the augmentation process naturally reduces to a standard form—ensuring adaptability across different radar data scenarios. 

Specifically, in our training, we apply class-aware oversampling such that samples from minority classes are duplicated and augmented more frequently, resulting in approximately 800 samples per class (six classes total) per epoch. The augmentation strategies are summarized as follows:

% ... (the rest of your subsection remains the same)

\begin{itemize}
    \item \textbf{Geometric Augmentations:} To simulate different target orientations and perspectives, we apply random rotations in increments of 90 degrees ($0^{\circ}, 90^{\circ}, 180^{\circ}, 270^{\circ}$) and random horizontal/vertical flips (50\% probability each).

    \item \textbf{Pixel-Level and Noise Augmentations:} 
    To improve the model's resilience to varying imaging conditions and sensor noise, we introduce several photometric distortions, including speckle noise inherent in SAR images, random gamma correction, Gaussian blurring, and adjustments to brightness/contrast.
    
    \item \textbf{1-bit Specific Augmentation:} 
    For the 1-bit images, we additionally implement a random erasing strategy. This technique randomly selects a rectangular region in an image and erases its pixels. This forces the model to learn from incomplete information and attend to a wider range of features, which is critical for the information-sparse 1-bit data.
\end{itemize}

All augmentations are applied randomly and exclusively during training. No augmentations are used during the validation or testing phases to ensure consistent performance evaluation.

%%%%%%%%%%%%%%%%%%%%%%%%%%%%%%%%%%%%%%%%%%%%%%%%%%%%%%%%%%%%%%%%%%%%%%%%%%%%%%%%
%                      SECTION IV: EXPERIMENTS (MERGED)
%%%%%%%%%%%%%%%%%%%%%%%%%%%%%%%%%%%%%%%%%%%%%%%%%%%%%%%%%%%%%%%%%%%%%%%%%%%%%%%%
\section{Experiments}
\label{sec:experiments}

In this section, we evaluate the performance of the proposed CF-Net.\footnote{Our source code is publicly available at \url{https://github.com/embedded-qjd/CF-Net}.} 
% To facilitate reproducibility and future research, the source code of the proposed CF-Net is made publicly available at \url{https://github.com/embedded-qjd/CF-Net}.
We first conduct comprehensive experiments on the SAR ship classification task to validate the core framework and perform ablation studies. 
Subsequently, to demonstrate the generalization capability of our method, we extend the evaluation to a Human Activity Recognition (HAR) task using millimeter-wave radar.

To quantitatively evaluate the classification performance, four commonly used metrics are adopted: 
Accuracy, Precision, Recall, and F1-score, which are computed from the confusion matrix components --- True Positives (TP), True Negatives (TN), False Positives (FP), and False Negatives (FN).
Accuracy measures the proportion of correctly classified samples, defined as $(\mathrm{TP + TN) / (TP + TN + FP + FN})$.
Precision evaluates the correctness of positive predictions, given by $\mathrm{TP / (TP + FP)}$.
Recall, or sensitivity, reflects the ability to identify all positive instances, calculated as $\mathrm{TP / (TP + FN)}$.
The F1-score provides a balanced harmonic mean of Precision and Recall, especially suitable for imbalanced datasets.
For the pre-training stage, we use the Peak Signal-to-Noise Ratio (PSNR) \cite{wang2004image} to measure the reconstruction quality of recovered images, where higher PSNR values indicate better fidelity.
This metric serves only as an auxiliary verification of the pre-training effectiveness and is not used in the final classification evaluation.

% =========================================================================
% SUBSECTION A: EXPERIMENT I - SAR Image Classification (Old Section IV)
% =========================================================================
\subsection{Experiment I: SAR Image Classification}
\label{sec:exp_sar}

\subsubsection{Dataset and Data Preprocessing}
\label{sec:dataset}

\begin{figure}[t!]
\centering
\subfloat[Bulk Carrier (Class 1)]{\includegraphics[width=0.28\linewidth]{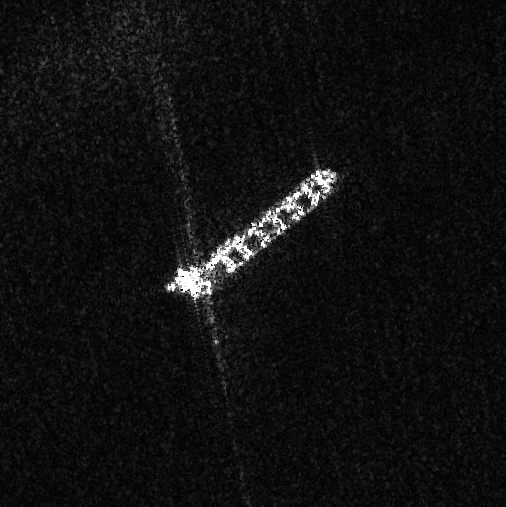}\label{fig:sample_bulkcarrier}}
\hfill
\subfloat[Dredger (Class 6)]{\includegraphics[width=0.28\linewidth]{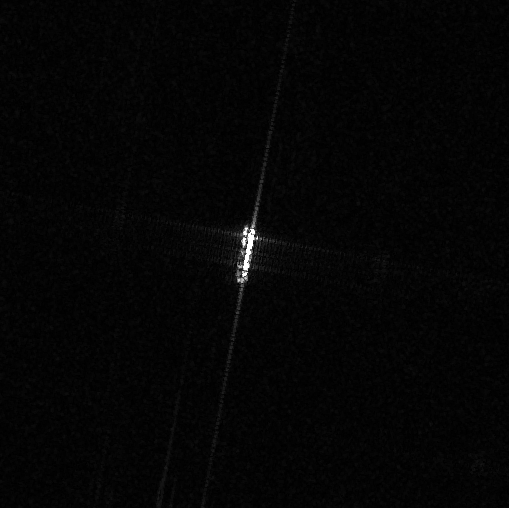}\label{fig:sample_dredger}}
\hfill
\subfloat[Fishing (Class 4)]{\includegraphics[width=0.28\linewidth]{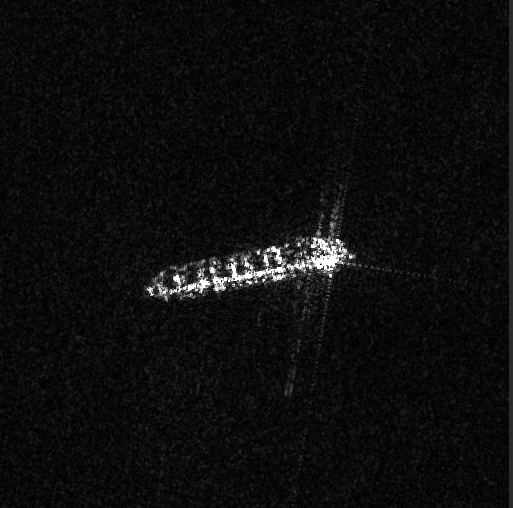}\label{fig:sample_fishing}}
\caption{Examples of SAR ship images used in the experiments.}
\label{fig:dataset_samples}
\end{figure}

\textbf{FUSAR-Ship Dataset:}
Experiments are conducted on the publicly available FUSAR-Ship dataset \cite{hou2020fusar}, constructed from imagery captured by China's Gaofen-3 (GF-3) C-band SAR satellite \cite{hou2020fusar}.
We select six distinct ship categories from the full dataset, partitioning it into training (70\%) and testing (30\%) sets, ensuring that samples from each class were distributed proportionally across both sets.
Detailed distribution is shown in Table~\ref{tab:dataset_distribution}, and representative visual examples of these categories are illustrated in Fig.~\ref{fig:dataset_samples}.

A major challenge with this dataset is the class imbalance as shown in Table~\ref{tab:dataset_distribution}, which can cause model bias and overfitting.
To address this, we employ the data augmentation strategies detailed in Section~\ref{sec:data_augmentation}.

\begin{figure}[t!]
\centering
\includegraphics[width=\columnwidth]{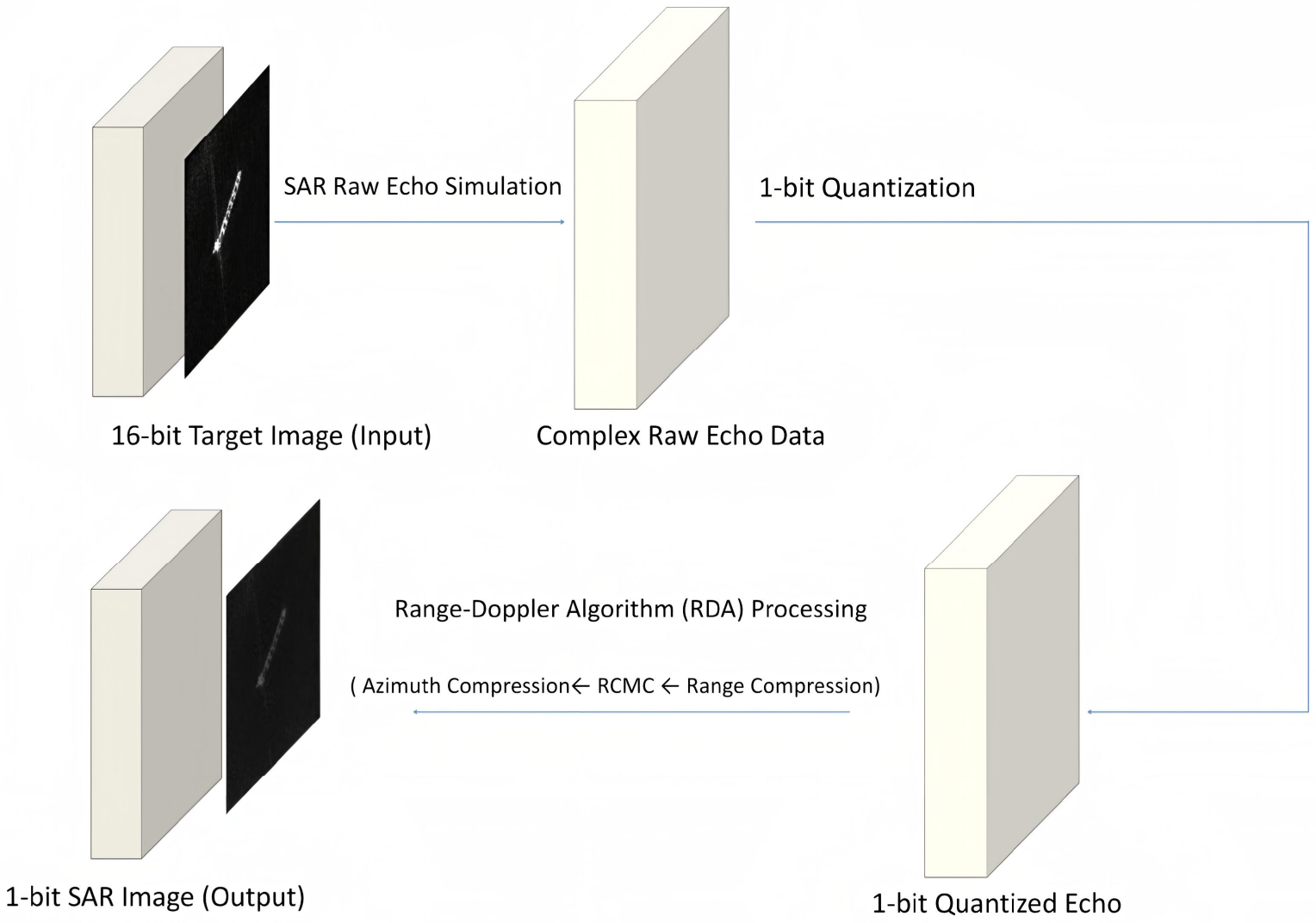}
\caption{The pipeline for generating 1-bit SAR images from 16-bit target images via Range-Doppler (RD) simulation.}
\label{fig:rd_pipeline}
\end{figure}

\textbf{1-bit Data Generation via RD Imaging:}
The original FUSAR-Ship dataset provides high-fidelity 16-bit SAR images (used as ground truth).
To generate the corresponding 1-bit dataset, we develop a simulation pipeline based on the classic Range-Doppler Algorithm (RDA), as illustrated in Fig.~\ref{fig:rd_pipeline}.
For each 16-bit target image, we treat it as a ground-truth reflectivity map.
First, a raw echo signal is simulated by modeling the SAR sensor's trajectory and pulse characteristics.
This complex-valued raw echo is then subjected to an extreme 1-bit quantization process.
Subsequently, this 1-bit quantized raw data is processed through a standard RDA chain—comprising range compression, Range Cell Migration Correction (RCMC), and azimuth compression—to form the final, information-sparse 1-bit SAR image.
This full process is summarized in Algorithm~\ref{alg:data_generation}.

% --- RD成像流程图 ---

% --- RD 成像伪代码 ---
\begin{algorithm}[t!]
\caption{1-bit SAR Image Generation via RD Simulation}
\label{alg:data_generation}
\begin{algorithmic}[1]
\State \textbf{Input:} 16-bit target image $I_{\mathrm{16bit}}$, Radar parameters $\mathcal{P}$
\State \textbf{Output:} 1-bit SAR image $I_{\mathrm{1bit}}$

\State // Step 1: Echo Simulation
\State Initialize reflectivity map $PSF \gets I_{\mathrm{16bit}}$
\State Initialize raw echo data $E_{\mathrm{raw}} \gets \mathrm{zeros}(N_{\mathrm{range}}, N_{\mathrm{azimuth}})$
\For{each azimuth position $k$}
    \For{each scene point $(jj, ii)$ in $PSF$}
        \State Calculate instantaneous slant range $R(k, jj, ii)$
        \State Generate point target echo $e_{\mathrm{point}}$ based on $R$ and $\mathcal{P}$
        \State $E_{\mathrm{raw}}[:, k] \gets E_{\mathrm{raw}}[:, k] + e_{\mathrm{point}} \times PSF(jj, ii)$
    \EndFor
\EndFor
\State
\State // Step 2: 1-bit Quantization
\State $E_{\mathrm{1bit,real}} \gets \mathrm{sign}(\mathrm{Real}(E_{\mathrm{raw}}))$
\State $E_{\mathrm{1bit,imag}} \gets \mathrm{sign}(\mathrm{Imag}(E_{\mathrm{raw}}))$
\State $E_{\mathrm{quantized}} \gets E_{\mathrm{1bit,real}} + \mathrm{i} \times E_{\mathrm{1bit,imag}}$
\State
\State // Step 3: Range-Doppler Algorithm (RDA) Processing
\State $E_{\mathrm{rc}} \gets \mathrm{RangeCompress}(E_{\mathrm{quantized}})$
\State $E_{\mathrm{rcmc}} \gets \mathrm{RCMC}(E_{\mathrm{rc}})$
\State $I_{\mathrm{1bit}} \gets \mathrm{AzimuthCompress}(E_{\mathrm{rcmc}})$
\State
\State \Return $I_{\mathrm{1bit}}$
\end{algorithmic}
\end{algorithm}

\begin{table}[t!]
\centering
\caption{Distribution of Ship Classes Used in Our Experiments (70/30 Split)}
\label{tab:dataset_distribution}
\begin{tabular}{|l|c|c|c|}
\hline
\textbf{Class Name} & \textbf{Training Set} & \textbf{Testing Set} & \textbf{Total} \\
\hline
\hline
Bulk Carrier      & 35  & 15  & 50    \\
Containership     & 35  & 15  & 50    \\
Tug               & 38  & 16  & 54    \\
Fishing           & 550 & 235 & 785   \\
Tanker            & 104 & 44  & 148   \\
Dredger           & 39  & 17  & 56    \\
\hline
\textbf{Total}    & \textbf{801} & \textbf{342} & \textbf{1143}  \\
\hline
\end{tabular}
\end{table}

% --- Algorithm 3 ---
\begin{algorithm}[htbp]
\caption{Overall CF-Net Framework}
\label{alg:overall_framework}
\begin{algorithmic}[1]
\State \textbf{Input:} Training data $\{ (X_{\mathrm{1bit}}, X_{\mathrm{16bit}}, Y) \}$
\State \textbf{Input:} Validation data $D_{\mathrm{val}}$
\State \textbf{Output:} Final fine-tuned classification model $\mathcal{G}_{\mathrm{final}}$

\Statex
\Statex \textbf{Stage 1: Self-Supervised Pre-training}
\State Initialize CF-Net model $\mathcal{F}$ (with $E_{\mathrm{1bit}}, E_{\mathrm{16bit}}$)
\State $E_{\mathrm{1bit}} \gets \mathrm{Pretrain}~\mathcal{F}~\mathrm{using}~(X_{\mathrm{1bit}}, X_{\mathrm{16bit}}, Y)$
\State \Comment{See Algorithm \ref{alg:pretraining} for pre-training details.}
\Statex

\Statex \textbf{Stage 2: Handcrafted Feature Extraction}
\State $\hat{X}_{\mathrm{16bit}} \gets \mathcal{F}_{\mathrm{S}}(X_{\mathrm{1bit}})$
\State $X_{\mathrm{HOG}} \gets \mathrm{ExtractHOGFeatures}(\hat{X}_{\mathrm{16bit}})$
\Statex

\Statex \textbf{Stage 3: Supervised Classification Fine-tuning}
\State Initialize classification model $\mathcal{G}$ with pre-trained $E_{\mathrm{1bit}}$
\State $\mathcal{G}_{\mathrm{final}} \gets \mathrm{Finetune}~\mathcal{G}~\mathrm{using}~(X_{\mathrm{1bit}}, X_{\mathrm{HOG}}, Y)$
\State \Comment{Input $X_{\mathrm{HOG}}$ is None if Stage 2 was skipped.}
\State \Comment{See Algorithm \ref{alg:finetuning} for fine-tuning details.}

\State \Return $\mathcal{G}_{\mathrm{final}}$
\end{algorithmic}
\end{algorithm}

\subsubsection{Implementation Details}
The overall procedure for SAR target classification experiments follows the framework outlined in Algorithm~\ref{alg:overall_framework}, encompassing pre-training, HOG feature extraction, and fine-tuning.
The framework is implemented using the PyTorch framework (version 2.6.0) with Python 3.13.
Experiments are conducted on a workstation equipped with an Intel Core i5-14600KF CPU, 32 GB of RAM, and an NVIDIA GeForce RTX 4060 GPU.
The operating system is Windows 11 Pro, and the models are accelerated using CUDA 12.6.
For the pre-training stage, the model is trained for 100 epochs using the AdamW optimizer with an initial learning rate of $1 \times 10^{-4}$.
The subsequent fine-tuning stage for classification is performed for 60 epochs, also with AdamW, and a learning rate of $5 \times 10^{-5}$.

\begin{figure}[t!]
\centering
\includegraphics[width=\columnwidth]{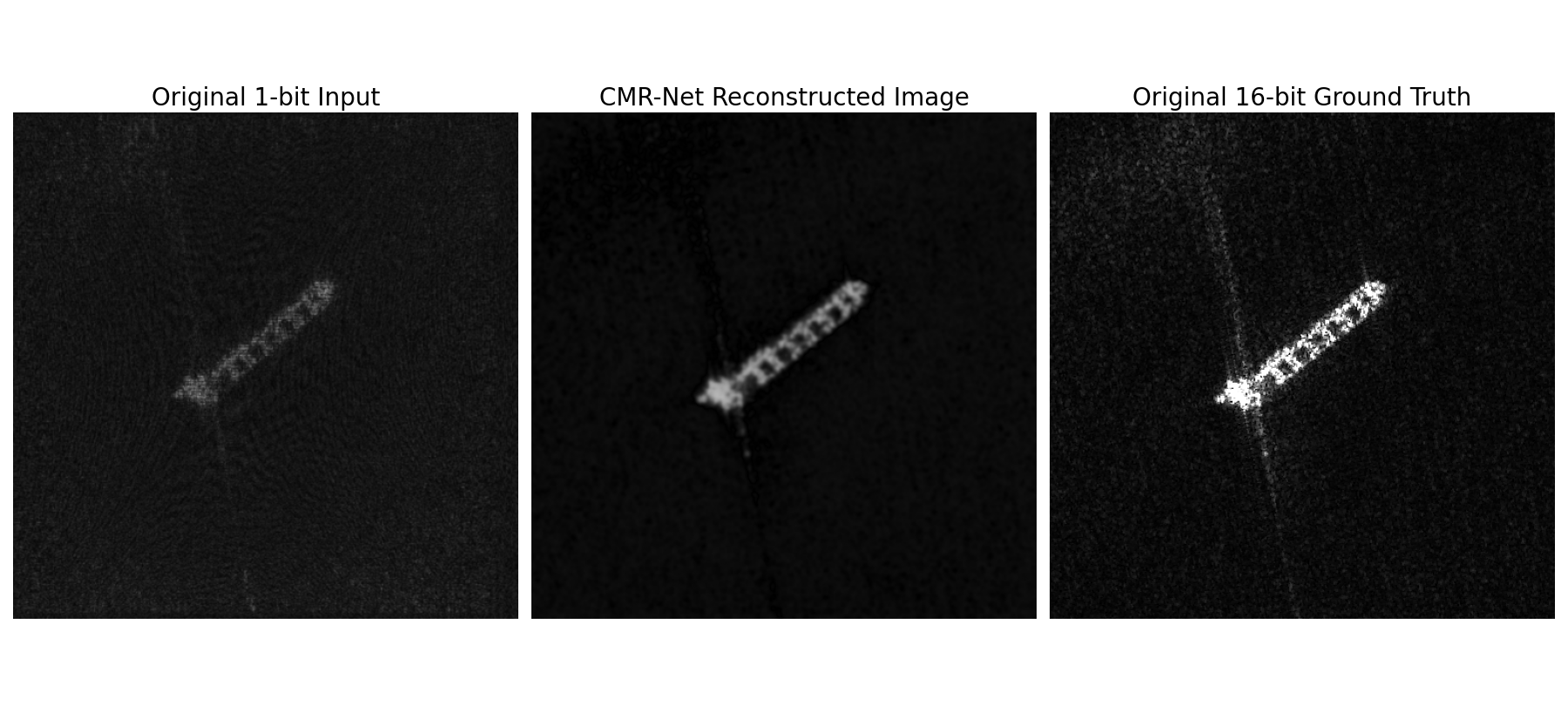}
\caption{Visual results of the cross-feature reconstruction stage.
For each target, we show the 1-bit input, the image reconstructed by our network, and the corresponding 16-bit ground truth.
The model demonstrates a strong capability to restore target structure and detail.}
\label{fig:reconstruction_results}
\end{figure}

\subsubsection{Evaluation of the Pre-training Stage}
\begin{figure}[t]
    \centering
    \includegraphics[width=\linewidth]{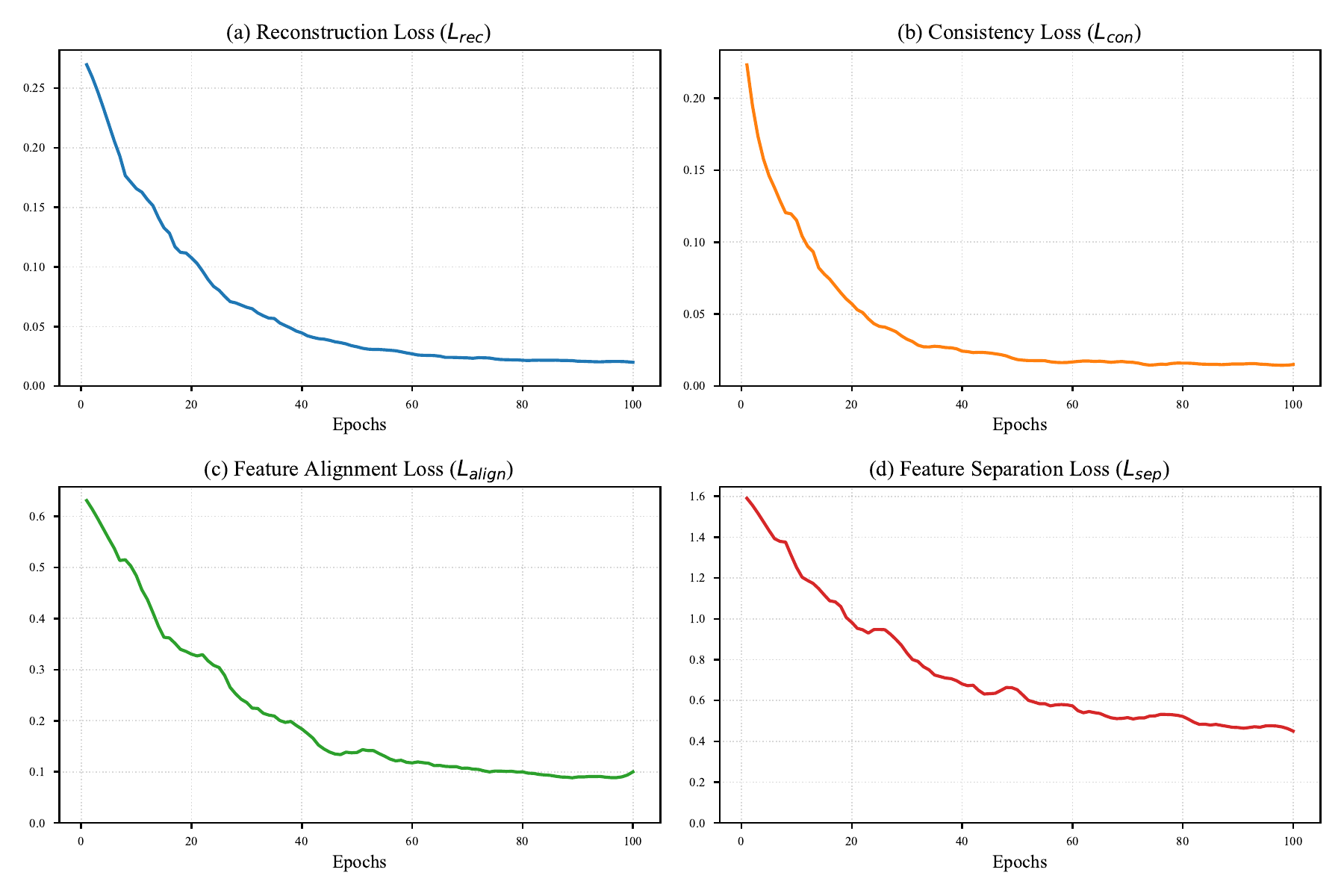} 
    \caption{Training convergence curves of the four component losses during the Stage 1 pre-training process. (a) Reconstruction Loss $L_{\mathrm{rec}}$. (b) Consistency Loss $L_{\mathrm{con}}$. (c) Feature Alignment Loss $L_{\mathrm{align}}$. (d) Feature Separation Loss $L_{\mathrm{sep}}$. All losses demonstrate stable convergence, validating the effectiveness of the proposed compound loss function.}
    \label{fig:stage1_losses}
\end{figure}

The foundational hypothesis of our framework is that the cross-feature reconstruction task enables the encoder to learn meaningful feature representations, which we validate through both quantitative and qualitative evaluations.
As a preliminary verification of the pre-training effectiveness, we first examine the reconstruction fidelity of the recovered images.
Quantitatively, CF-Net achieves a PSNR of 25.24~dB on the test set when reconstructing 16-bit images from 1-bit inputs, demonstrating good recovery quality \cite{wang2004image}.
Qualitatively, as shown in Fig.~\ref{fig:reconstruction_results}, the network successfully restores key structural details and suppresses severe noise in 1-bit SAR images, with reconstructed results closely resembling ground-truth 16-bit images.
These observations confirm that the encoder effectively learns essential target signatures despite extreme information loss, providing a solid foundation for subsequent fine-tuning and classification. To further verify the stability of the optimization process, the training curves of the four component losses are illustrated in Fig.~\ref{fig:stage1_losses}. It can be observed that all loss terms decrease steadily and converge to a stable state, indicating that the encoder successfully learns robust feature representations through the cross-feature reconstruction task.

\subsubsection{Baseline Comparison with Single-Scale Fusion}
\label{sec:single_scale_comparison}
To comprehensively evaluate our framework, we first establish strong baselines using a single-scale feature fusion architecture, where only the deepest encoder layer’s feature map is used for classification.
This relying exclusively on high-level semantic features. Specifically, the encoder's output feature map is processed via global average pooling layer to form a feature vector, concatenated with a precomputed HOG vector, and fed into a classification head.
we compare three single-scale models with identical architectures but different training settings.
The first model is a baseline ResNet-34 trained directly on 1-bit images, where HOG features were extracted from the same 1-bit data and fused at the classifier level.
The second uses our proposed pre-training approach: the encoder was first pre-trained using the cross-feature reconstruction strategy and then fine-tuned on 1-bit classification, also fused with HOG features.
For reference, we additionally train the same architecture on 16-bit images to indicate the performance achievable with full-precision inputs.
As summarized in Table~\ref{tab:single_scale_results}, the baseline trained directly on 1-bit data achieves only 66.28\% accuracy, while our pre-trained single-scale model reaches 76.74\%, yielding over 10 percentage points of improvement.
In comparison, the same model trained on 16-bit data attains 80.00\%, showing that our pre-training method effectively transfers high-level structural priors and recovers most of the discriminative capacity lost due to 1-bit quantization.
This finding also motivates the introduction of the multi-scale fusion strategy to further strengthen low-level feature utilization.

\begin{table}[t!]
\centering
\caption{Performance Comparison of Single-Scale Fusion Models on the Test Set.}
\label{tab:single_scale_results}
\begin{tabular}{|l|c|c|}
\hline
\textbf{Method} & \textbf{Input Data} & \textbf{Accuracy (\%)} \\
\hline
\hline
1-bit ResNet + HOG (Baseline) & 1-bit & 66.28 \\
\textbf{CF-Net (Single-Scale) + HOG} & \textbf{1-bit} & \textbf{76.74} \\
\hline
16-bit ResNet + HOG & 16-bit & 80.00 \\
\hline
\end{tabular}
\end{table}

The matrix reveals that our model performs exceptionally well on the majority class, Fishing (C4), correctly identifying 108 out of 118 samples.
However, it exhibits some confusion among the minority classes with fewer samples.
For instance, Dredger (C6) and Tug (C3) are the most challenging categories, often being misclassified as other vessel types.
This detailed analysis indicates that while the overall performance is strong, future work could focus on improving feature discrimination for these rarer ship classes.

\begin{table*}[t!]
\centering
\renewcommand{\arraystretch}{1.3} 
\setlength{\tabcolsep}{10pt}
\caption{Performance Comparison with Fine-tuned Baseline Models on the FUSAR-Ship Dataset.}
\label{tab:main_results}
\begin{tabular}{|l|c|c|c|}
\hline
\textbf{Method} & \textbf{Input Data} & \textbf{Accuracy (\%)} & \textbf{F1-score (\%)} \\
\hline
\hline
\multicolumn{4}{|l|}{\textit{Pre-trained Backbones Fine-tuned on 16-bit Data}} \\
\hline
ResNet-18 (ImageNet Pre-trained)    & 16-bit & 77.29 & 46.47 \\
ResNet-34 (ImageNet Pre-trained)    & 16-bit & 77.73 & 43.69 \\
ResNet-50 (ImageNet Pre-trained)    & 16-bit & 79.04 & 47.93 \\
VGG-16 (ImageNet Pre-trained)       & 16-bit & 77.73 & 46.58 \\
VGG-19 (ImageNet Pre-trained)       & 16-bit & 78.60 & 39.43 \\
DenseNet-121 (ImageNet Pre-trained) & 16-bit & 75.98 & 37.93 \\
DenseNet-169 (ImageNet Pre-trained) & 16-bit & 78.17 & 46.77 \\
\hline
\hline
\multicolumn{4}{|l|}{\textit{Pre-trained Backbones Fine-tuned on 1-bit Data}} \\
\hline
ResNet-18 (ImageNet Pre-trained)    & 1-bit & 70.41 & 22.63 \\
ResNet-34 (ImageNet Pre-trained)    & 1-bit & 69.84 & 16.62 \\
ResNet-50 (ImageNet Pre-trained)    & 1-bit & 68.37 & 17.32 \\ 
VGG-16 (ImageNet Pre-trained)       & 1-bit & 70.12 & 28.93 \\
VGG-19 (ImageNet Pre-trained)       & 1-bit & 69.27 & 20.74 \\
DenseNet-121 (ImageNet Pre-trained) & 1-bit & 67.58 & 13.56 \\
DenseNet-169 (ImageNet Pre-trained) & 1-bit & 71.03 & 16.41 \\
\hline
\hline
\multicolumn{4}{|l|}{\textit{Proposed Method}} \\
\hline
\textbf{CF-Net (Ours)} & \textbf{1-bit} & \textbf{80.81} & \textbf{56.58} \\
\hline
\end{tabular}
\end{table*}

\begin{table}[t!]
\centering
\caption{Confusion Matrix of the Proposed CF-Net on the Test Set.
C1: Bulk Carrier, C2: Containership, C3: Tug, C4: Fishing, C5: Tanker, C6: Dredger.}
\label{tab:confusion_matrix}
\setlength{\tabcolsep}{4pt} 
\renewcommand{\arraystretch}{1.2} 
\begin{tabular}{c|c|cccccc|c}
\hline
\hline
\multicolumn{2}{c|}{\multirow{2}{*}{\textbf{}}} & \multicolumn{6}{c|}{\textbf{Predicted Class}} & \\
\cline{3-8}
\multicolumn{2}{c|}{} & C1 & C2 & C3 & C4 & C5 & C6 & \textbf{Total} \\
\hline
\multirow{6}{*}{\rotatebox{90}{\textbf{Actual Class}}}
& \textbf{C1} & \textbf{3} & 1 & 2 & 1 & 1 & 0 & 8 \\
& \textbf{C2} & 0 & \textbf{5} & 0 & 1 & 1 & 0 & 7 \\
& \textbf{C3} & 1 & 1 & \textbf{3} & 2 & 1 & 0 & 8 \\
& \textbf{C4} & 0 & 3 & 2 & \textbf{108} & 4 & 1 & 118 \\
& \textbf{C5} & 0 & 2 & 0 & 2 & \textbf{18} & 0 & 22 \\
& \textbf{C6} & 0 & 0 & 0 & 4 & 3 & \textbf{2} & 9 \\
\hline
\multicolumn{2}{c|}{\textbf{Total (Predicted)}} & 4 & 12 & 7 & 118 & 28 & 3 & \textbf{172} \\
\hline
\hline
\end{tabular}
\end{table}

\subsubsection{Ablation Study}
\begin{figure}[htbp]
    \centering
    \includegraphics[width=0.8\linewidth]{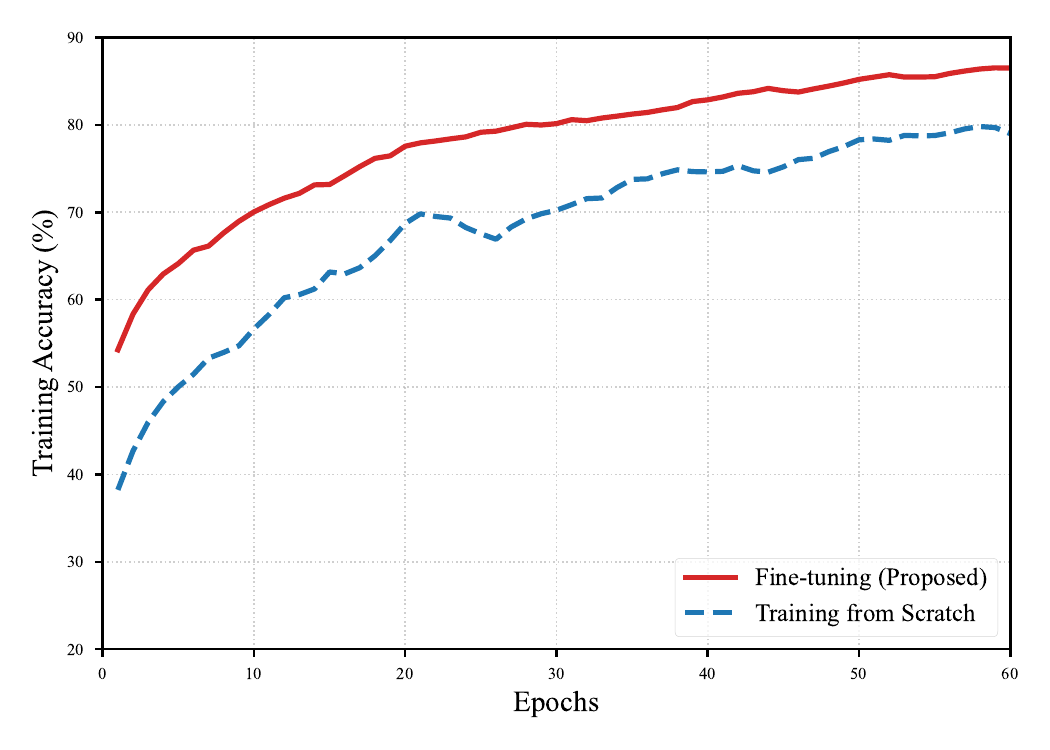}
    \caption{Comparison of training accuracy curves between the proposed Fine-tuning strategy and Training from Scratch. The proposed method demonstrates a higher starting point and faster convergence.}
    \label{fig:stage2_acc}
\end{figure}

\begin{figure}[htbp]
    \centering
    \includegraphics[width=0.8\linewidth]{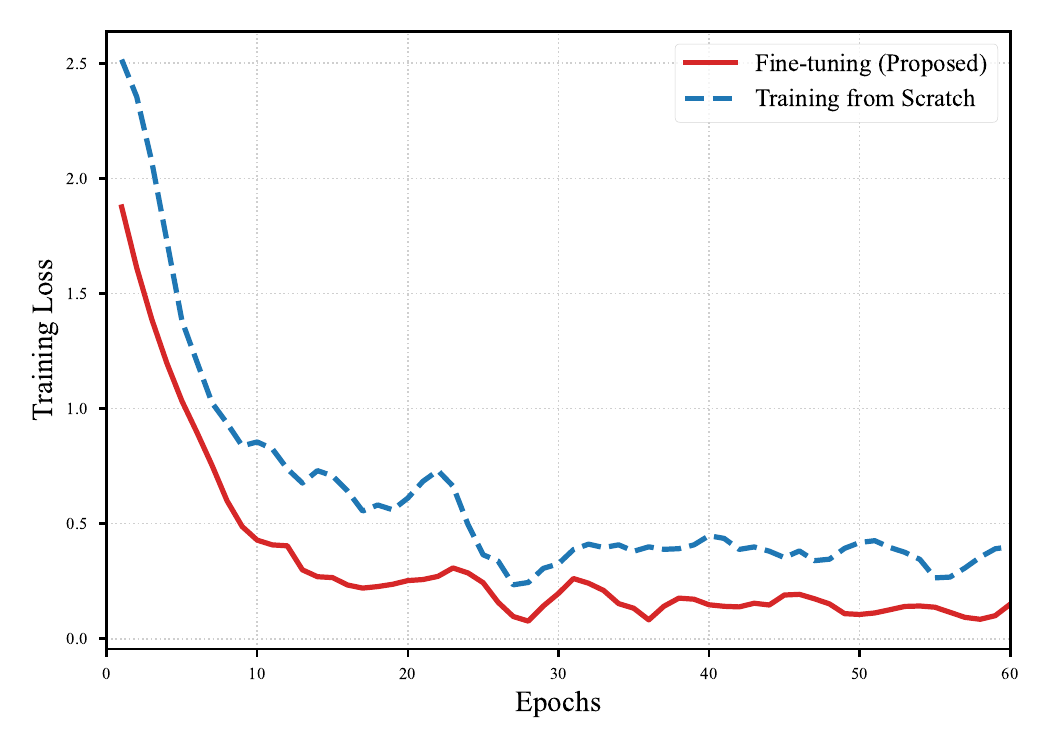}
    \caption{Comparison of training loss convergence between the proposed Fine-tuning strategy and Training from Scratch.}
    \label{fig:stage2_loss}
\end{figure}

To validate the effectiveness of key components in CF-Net, we conduct a series of ablation studies:

\paragraph{Fine-tuning vs. Training from Scratch}
A fundamental question in model training is whether to fine-tune pre-trained weights or train the network entirely from scratch. We compare these two strategies for our multi-scale fusion network using 1-bit data.
In the fine-tuning setup, the ResNet-34 backbone is initialized with ImageNet pre-trained weights and then trained end-to-end, while the from-scratch variant starts from random initialization.
As shown in Table~\ref{tab:ablation_finetune_vs_scratch}, fine-tuning yields a clear performance advantage, achieving 80.81\% accuracy and 56.58\% F1-score compared with 74.23\% and 44.37\% when trained from scratch.
The large margin highlights the importance of transfer learning, as pre-trained weights provide stable low-level feature representations that help the model converge faster and generalize better under extremely quantized 1-bit conditions. To provide a more intuitive comparison of the training dynamics, the training accuracy and loss curves are visualized in Fig.~\ref{fig:stage2_acc} and Fig.~\ref{fig:stage2_loss}, respectively. As shown in Fig.~\ref{fig:stage2_acc}, the proposed fine-tuning strategy (red line) exhibits a significantly higher starting accuracy and a faster convergence rate compared to training from scratch (blue line), thanks to the structural priors learned during the pre-training stage. Similarly, Fig.~\ref{fig:stage2_loss} confirms that our method achieves a lower training loss with more stable gradients.
Therefore, the fine-tuning strategy is adopted for all subsequent experiments in this work.
% === Stage 2 Accuracy Figure ===

% === Stage 2 Loss Figure ===
\begin{table}[t!]
\centering
\caption{Comparison between Fine-tuning and Training from Scratch on 1-bit Data.}
\label{tab:ablation_finetune_vs_scratch}
\renewcommand{\arraystretch}{1.35}
\setlength{\tabcolsep}{6pt}
\begin{tabular}{|l|c|c|}
\hline
\textbf{Training Strategy} & \textbf{Accuracy (\%)} & \textbf{F1-score (\%)} \\
\hline
\hline
\textbf{Fine-tuning (Proposed)} & \textbf{80.81} & \textbf{56.58} \\
Training from Scratch           & 74.23          & 44.37 \\
\hline
\end{tabular}
\end{table}

\paragraph{Effectiveness of Pre-training and Multi-Scale Feature Fusion}
We conduct a unified ablation study to investigate the roles of the pre-training stage and the multi-scale fusion strategy within the proposed framework.
The results are summarized in Table~\ref{tab:ablation_pretrain_multiscale}.  

Starting from a baseline trained directly on 1-bit data, the model achieves 68.70\% accuracy and 31.54\% F1-score.
Incorporating the cross-feature pre-training stage provides a large boost, improving accuracy to 76.74\% and F1-score to 50.38\%.
This demonstrates that pre-training effectively transfers structural priors learned from high-quality data, allowing the network to better interpret severely quantized inputs.
Building upon the pre-trained backbone, we further introduce multi-scale feature fusion by integrating features from different encoder depths.
As the number of scales increases, both accuracy and F1-score steadily improve, ultimately reaching 80.81\% and 56.58\% with the full five-scale configuration.
This trend confirms that multi-scale aggregation enriches the feature representation by combining high-level semantic information with low-level structural cues, which are essential for distinguishing visually similar ship classes under 1-bit quantization.
Overall, the pre-training and multi-scale modules complement each other: pre-training enhances feature quality, while multi-scale fusion strengthens feature diversity and robustness, jointly leading to the best overall performance.

\begin{table}[t!]
\centering
\caption{Ablation Study on the Effectiveness of Pre-training and Multi-Scale Feature Fusion.
Results are reported on the test set.}
\label{tab:ablation_pretrain_multiscale}
\renewcommand{\arraystretch}{1.5}
\setlength{\tabcolsep}{6.5pt}
\begin{tabular}{|l|c|c|}
\hline
\textbf{Configuration} & \textbf{Accuracy (\%)} & \textbf{F1-score (\%)} \\
\hline
\hline
Baseline (1-bit, no pre-training) & 68.70 & 31.54 \\
\hline
\textbf{+ Pre-training (Single-Scale)} & \textbf{76.74} & \textbf{50.38} \\
\hline
\multicolumn{3}{|l|}{\textit{+ Multi-Scale Feature Fusion (after Pre-training):}} \\
\hline
2 Scales & 78.09 & 52.27 \\
3 Scales & 79.42 & 54.36 \\
4 Scales & 80.37 & 55.92 \\
5 (Full Multi-Scale) & \textbf{80.81} & \textbf{56.58} \\
\hline
\end{tabular}
\end{table}

\paragraph{Analysis of Loss Function Components}
We analyze the contribution of each component in the compound pre-training loss.
Results in Table~\ref{tab:ablation_loss} show that both feature alignment ($L_{\mathrm{align}}$) and feature separation ($L_{\mathrm{sep}}$) losses provide significant performance gains over a baseline with only reconstruction loss.
Notably, removing the separation loss results in the largest performance drop, highlighting its importance in structuring the feature space for classification.

\begin{table}[t!]
\centering
\caption{Ablation Study on the Components of the Compound Loss.}
\label{tab:ablation_loss}
\renewcommand{\arraystretch}{1.3}
\begin{tabular}{|l|c|}
\hline
\textbf{Loss Configuration} & \textbf{Accuracy (\%)} \\
\hline
\hline
$L_{\mathrm{rec}}$ only & 72.51 \\
$L_{\mathrm{rec}} + L_{\mathrm{con}} + L_{\mathrm{align}}$ & 78.65 \\
\textbf{Full Loss (All Components)} & \textbf{80.81} \\
\hline
\end{tabular}
\end{table}

\paragraph{Effect of HOG Feature Source on Model Performance}
To examine how the source of handcrafted features affects model performance, we conduct two complementary experiments:  
(1) using HOG features as auxiliary inputs in our full fusion model, and  
(2) using only HOG features to train a lightweight MLP classifier that directly reflects their intrinsic quality.
As shown in Table~\ref{tab:ablation_hog_combined}, when HOG features are extracted from the reconstructed images, the full fusion network achieves 80.81\% accuracy, whereas using HOG features directly from raw 1-bit data drastically reduces accuracy to 44.40\%.
A similar improvement is observed in the HOG-only classifier, where features from reconstructed images reach 70.57\%, far outperforming those from 1-bit inputs (43.50\%).
These results demonstrate that the reconstruction network effectively restores structural and gradient information that is severely distorted in raw 1-bit measurements.
Consequently, the CF-Net not only benefits the end-to-end fusion process but also substantially enhances the discriminative power of handcrafted features such as HOG.

\begin{table}[t!]
\centering
\caption{Ablation Study on the Impact of HOG Feature Source on Model Performance.}
\label{tab:ablation_hog_combined}
\renewcommand{\arraystretch}{1.3}
\setlength{\tabcolsep}{5.2pt} 
\begin{tabular}{|p{3.4cm}|c|c|}
\hline
\textbf{HOG Feature Source} & \textbf{Fusion Model (\%)} & \textbf{HOG-Only (\%)} \\
\hline
\hline
\textbf{From Reconstructed Image (Ours)} & \textbf{80.81} & \textbf{70.57} \\
From Raw 1-bit Image & 44.40 & 43.50 \\
\hline
\end{tabular}
\end{table}

\paragraph{Robustness to Varying Numbers of Classes}
To evaluate the robustness and scalability of our proposed framework under varying task complexities, we extend our evaluation to scenarios with different numbers of target classes.
Based on our core 6-class dataset, we construct additional 4-class and 5-class tasks by systematically removing the classes with the fewest samples.
Furthermore, to create a more challenging 7-class scenario, we augment the core set with the next most populous available class, ``Cargo". It is critical to note that the ``Cargo" class is substantially larger than the others (1,693 samples), introducing severe class imbalance into the 7-class task.
This makes the 7-class experiment a stringent stress test of a model's ability to handle both increased class diversity and data imbalance simultaneously.
For a fair comparison, all experiments in this section utilize the same network architecture and training configurations, with all models operating exclusively on 1-bit data.
We compare our proposed CF-Net against two strong baselines, ResNet-50 and HOG-ShipCLSNet, on the 4-class and 7-class tasks to gauge relative performance.
The results are summarized in Table~\ref{tab:varying_classes}.

As anticipated, our model's performance shows a general decline as the number of classes increases, reflecting the escalating task difficulty.
Crucially, our proposed CF-Net consistently and significantly outperforms both baselines in the 4-class and 7-class comparisons.
The performance gap is particularly pronounced in the challenging 7-class case.
While the baseline models suffer a steep performance drop when faced with the dual challenge of more classes and severe imbalance, our method maintains a much more graceful degradation.
This strongly suggests that the features learned via our cross-feature pre-training are inherently more robust and discriminative, enabling the model to scale more effectively to complex, real-world classification problems.

\begin{table}[t!]
\centering
\caption{Performance on Tasks with Varying Numbers of Classes (All on 1-bit Input).}
\label{tab:varying_classes}
\renewcommand{\arraystretch}{1.5}
\begin{tabular}{|c|l|c|c|}
\hline
\textbf{\# of Classes} & \textbf{Method} & \textbf{Accuracy (\%)} & \textbf{F1-score (\%)} \\
\hline
\hline
\multirow{3}{*}{4-Class} 
& \textbf{CF-Net (Ours)} & \textbf{83.51} & \textbf{64.77} \\
& HOG-ShipCLSNet          & 81.27          & 56.66 \\
& ResNet-50               & 79.89          & 53.31 \\
\hline
\multirow{3}{*}{5-Class} 
& \textbf{CF-Net (Ours)} & \textbf{82.73} & \textbf{57.41} \\
& HOG-ShipCLSNet          & 78.64          & 49.83 \\
& ResNet-50               & 73.92          & 39.78 \\
\hline
\multirow{3}{*}{6-Class} 
& \textbf{CF-Net (Ours)} & \textbf{80.81} & \textbf{56.58} \\
& HOG-ShipCLSNet          & 75.12          & 45.37 \\
& ResNet-50               & 68.37          & 17.32 \\ 
\hline
\multirow{3}{*}{7-Class} 
& \textbf{CF-Net (Ours)} & \textbf{75.07} & \textbf{50.25} \\
& HOG-ShipCLSNet          & 71.80          & 42.36 \\
& ResNet-50               & 64.90          & 15.27 \\
\hline
\end{tabular}
\end{table}

\subsubsection{Main Results}
To contextualize the performance of our framework, we compare it against widely used deep learning architectures including ResNet, VGG, DenseNet families.
Each model is initialized with ImageNet pre-trained weights and fine-tuned under two conditions: (1) full-precision 16-bit data (to establish benchmarks) and (2) 1-bit data (to serve as direct baselines for our method).
Comprehensive results are presented in Table~\ref{tab:main_results}. 

Baseline models on 1-bit data struggle to exceed 70\% accuracy—highlighting the task’s inherent difficulty.
In stark contrast, the proposed CF-Net achieves a final classification accuracy of 80.81\% using only the extremely compressed 1-bit data.
This result is remarkable: it not only outperforms all baseline models operating on 1-bit input but also provides 2\% to 5\% performance improvement compared to the methods rely on full-precision 16-bit data \cite{zhang2022hog}.
This establishes a strong new benchmark for 1-bit SAR classification. Furthermore, the detailed confusion matrix result is presented in Table~\ref{tab:confusion_matrix}.

% =========================================================================
% SUBSECTION B: EXPERIMENT II - Generalization to HAR
% =========================================================================
\subsection{Experiment II: Generalization to 1-bit Radar Human Activity Recognition}
\label{sec:har_validation}

To evaluate the versatility and generalizability of our proposed CF-Net framework, we apply its core two-stage learning paradigm to a different and challenging task: human activity recognition (HAR) using a self-collected millimeter-wave radar dataset (publicly available at \url{https://github.com/embedded-qjd/HAR-Dataset-Project}).

\subsubsection{HAR Dataset and Pre-processing}
\begin{figure}[htbp]
    \centering
    \includegraphics[width=\linewidth]{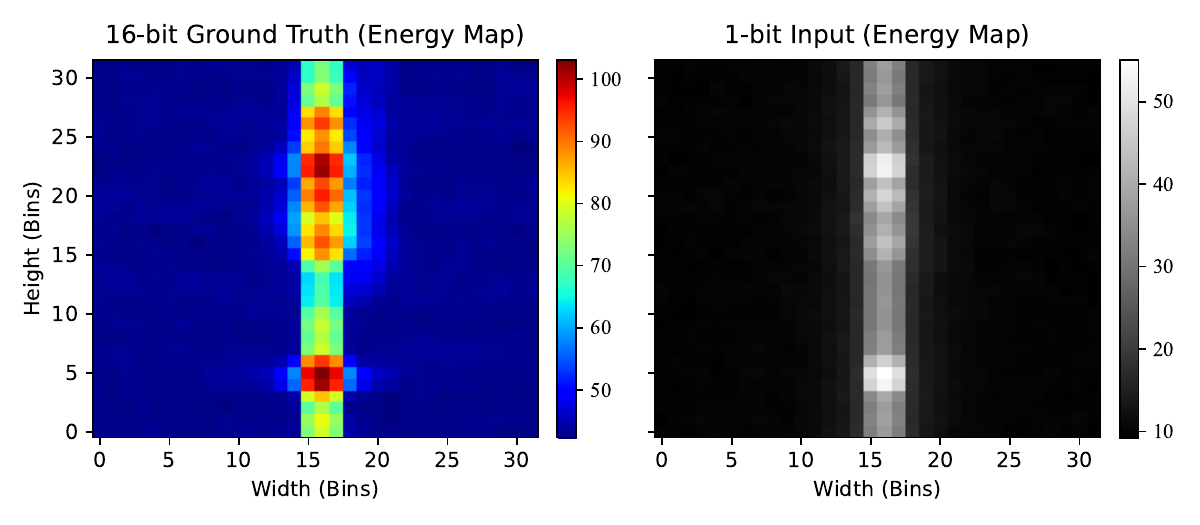} 
    \caption{Visual comparison of the aggregated energy maps for a sample human activity. Left: High-fidelity 16-bit ground truth. Right: 1-bit quantized input used for classification. The aggregation process preserves the dominant spatial structure despite the extreme quantization.}
    \label{fig:har_energy_map}
\end{figure}

The dataset is collected using an AWR2243 millimeter-wave radar sensor and comprises 10 distinct human activities (e.g., waving, kicking, and sitting down).
To ensure diversity, the data is recorded across three different environments: an indoor office, a narrow corridor, and a spacious outdoor plaza.
Each raw data sample is a 4D spatiotemporal radar cube, with dimensions representing channels, time-frames, height, and width ($X \in \mathbb{R}^{C \times T \times H \times W}$), where C=4, T=128, H=32, and W=32 in our case.
To adapt this rich, dynamic data for our 2D CNN-based pipeline, we project the 4D cube onto a static 2D plane by performing mean aggregation across both the channel (axis=0) and time (axis=1) dimensions.
This operation generates a single-channel aggregated map $X'$ that encapsulates the overall energy distribution of the action over its entire duration.
For any spatial coordinate $(h, w)$, the value of the aggregated map is computed as:
\begin{equation}
\label{eq:har_aggregation}
X'_{h,w} = \frac{1}{C \times T} \sum_{c=1}^{C} \sum_{t=1}^{T} X_{c,t,h,w}
\end{equation}
This pre-processing step is a critical adaptation that allows us to leverage our 2D image reconstruction and classification framework for spatiotemporal radar data. A visual comparison between the high-fidelity 16-bit ground truth and the 1-bit quantized aggregated map is presented in Fig.~\ref{fig:har_energy_map}.

% === 插入 HAR 能量图对比 ===
% ==========================

\subsubsection{CF-Net Adaptation for HAR}
\begin{figure}[t!]
\centering
\includegraphics[width=1.0\columnwidth]{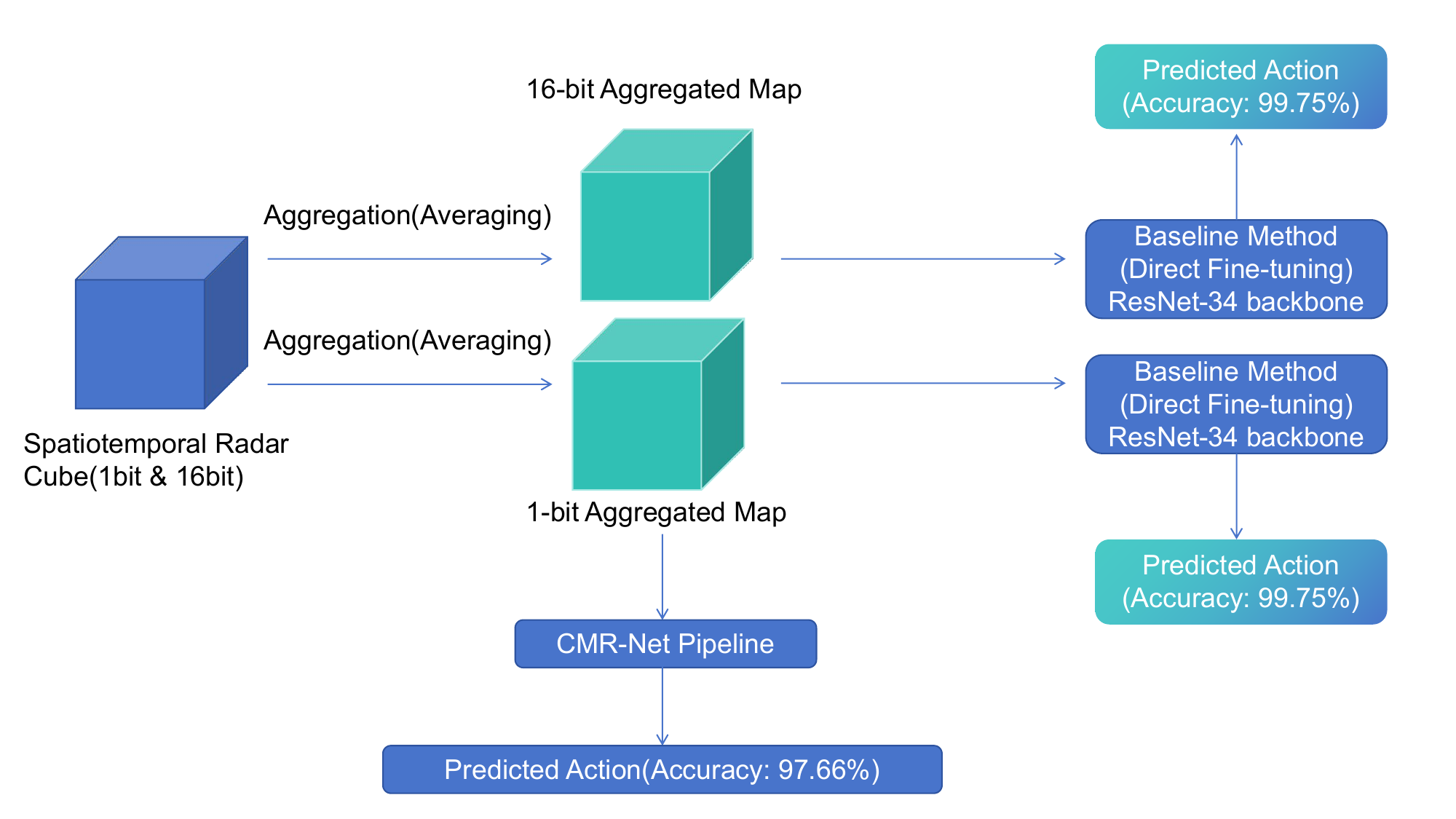}

\caption{The adapted CF-Net pipeline for the HAR task. The 1-bit and 16-bit aggregated maps are processed by baseline methods for performance bounds.
Our proposed method first uses the CF-Net recovery network (Stage 1) to restore a high-quality image from the 1-bit map, which is then fed into a dedicated classifier (Stage 2), yielding significantly improved accuracy.}
\label{fig:har_pipeline}
\end{figure}

For the HAR task, we adapt our framework into a sequential two-stage pipeline, as illustrated in Fig.~\ref{fig:har_pipeline}.
The pipeline is designed to first recover a high-fidelity representation from the 1-bit aggregated map and then perform classification.

The core components and adaptations of the pipeline are:
\begin{itemize}
    \item \textbf{Stage 1: Image Reconstruction.} The first stage utilizes the same architecture from our SAR experiments, pre-trained on the HAR dataset for the cross-feature task of reconstructing 16-bit aggregated maps from their 1-bit counterparts.
    During the final classification pipeline, this network's weights are frozen, and it functions purely as a high-fidelity image generator.
    \item \textbf{Stage 2: Classification.} The second stage employs a fine-tuned classifier that takes the reconstructed image from Stage 1 as input.
    A key adaptation is made to the classifier's architecture. Instead of the deep, multi-scale network used for SAR, we utilize a shallower feature extractor consisting of only the first three stages of a pre-trained ResNet-34.
    This modification is a deliberate design choice to prevent overfitting.
    The aggregated 2D maps, while informative, are less complex in texture and detail than SAR ship images.
    The shallower features are sufficient to capture the discriminative spatial patterns, whereas deeper features might lead to memorizing non-essential details.
    \item \textbf{Omission of HOG Features.} Unlike the SAR classification network, HOG feature fusion is intentionally omitted in this pipeline, deviating from the general framework presented in Algorithm~\ref{alg:overall_framework}.
    Specifically, the handcrafted feature extraction stage (lines 10-11 in Algorithm~\ref{alg:overall_framework}) is skipped.
    Preliminary experiments reveal that HOG features, which are designed for static image textures, did not improve performance when applied to the aggregated spatiotemporal energy maps.
    We hypothesize that the aggregation process already captures the dominant spatial information required for classification, making traditional texture descriptors redundant or even counterproductive for this data type.
    The subsequent fine-tuning stage (Stage 3 in Algorithm~\ref{alg:overall_framework}) therefore uses only the deep features from the pre-trained encoder.
\end{itemize}
\begin{table*}[t!]
\centering
\renewcommand{\arraystretch}{1.2}
\setlength{\tabcolsep}{12pt}
\caption{Performance Comparison on the HAR Dataset for 1-bit and 16-bit Inputs.}
\label{tab:har_results}
\begin{tabular}{|l|c|c|}
\hline
\textbf{Method} & \textbf{Input Data} & \textbf{Accuracy (\%)} \\
\hline
\hline
\multicolumn{3}{|l|}{\textit{\textbf{Core Pipeline Comparison}}} \\
\hline
\textbf{CF-Net Pipeline (Ours)} & \textbf{1-bit} & \textbf{97.66} \\
Direct Baseline      & 1-bit          & 81.27 \\
\hline
\hline
\multicolumn{3}{|l|}{\textit{Other Baseline Architectures}} \\
\hline
\multirow{2}{*}{ResNet-18} & 1-bit & 90.22 \\
                           & 16-bit & 96.49 \\
\hline
\multirow{2}{*}{ResNet-34} & 1-bit & 89.13 \\
                           & 16-bit & 97.41 \\
\hline
\multirow{2}{*}{ResNet-50} & 1-bit & 90.47 \\
                           & 16-bit & 97.58 \\
\hline
\multirow{2}{*}{ResNet-101} & 1-bit & 89.63 \\
                            & 16-bit & 98.83 \\
\hline
\multirow{2}{*}{VGG-16} & 1-bit & 90.38 \\
                        & 16-bit & 97.16 \\
\hline
\multirow{2}{*}{VGG-19} & 1-bit & 89.30 \\
                        & 16-bit & 97.32 \\
\hline
\multirow{2}{*}{DenseNet-121} & 1-bit & 91.39 \\
                              & 16-bit & 98.33 \\
\hline
\multirow{2}{*}{DenseNet-169} & 1-bit & 90.30 \\
                              & 16-bit & 98.66 \\
\hline
\end{tabular}
\end{table*}
\subsubsection{Results and Discussion}
To provide a comprehensive evaluation, we establish direct performance bounds for our pipeline using a standard ResNet-34 backbone, fine-tuned on both 1-bit and 16-bit aggregated maps.
We further compare our method against a wide array of other deep learning architectures.
The comprehensive results are presented in Table~\ref{tab:har_results}. The direct baseline achieves 81.27\% accuracy on 1-bit data, our proposed CF-Net pipeline, despite operating on the same challenging 1-bit input, achieves a remarkable accuracy of \textbf{97.66\%}.
This result powerfully validates our framework's ability to restore critical information lost during extreme quantization.
Furthermore, our method significantly outperforms all other baseline architectures on the 1-bit data, and its performance is competitive with, or even surpasses, many of these baselines when they operate on full-precision 16-bit data.
This success on a completely different data domain and task provides powerful evidence for the generalizability and effectiveness of the CF-Net framework.

\section{Conclusion}
\label{sec:conclusion}

This paper presented CF-Net, a novel and general framework for high-accuracy target classification from extremely quantized 1-bit radar data. Unlike prior approaches limited to isolated data representations, our method leverages cross-feature reconstruction and knowledge distillation to recover rich semantic information lost during 1-bit quantization. Through a self-supervised pre-training stage that reconstructs 16-bit images from 1-bit inputs using a compound loss, the encoder learns highly robust and discriminative representations, which are subsequently fine-tuned for efficient classification.

Comprehensive experiments on two distinct radar tasks, SAR ship classification and HAR, demonstrate the versatility and strong generalization ability of the proposed framework. CF-Net achieves 80.81\% accuracy on the 1-bit FUSAR-Ship dataset and 97.66\% accuracy on the 1-bit HAR dataset. These results establish a new state-of-the-art benchmark for 1-bit radar classification and confirm the feasibility of building high-performance radar perception systems directly from extremely low-bit measurements.

In future work, we plan to further compress the classification network toward a fully hardware-efficient implementation and explore lightweight feature fusion strategies to enhance real-time performance on embedded radar platforms. Overall, this study pioneers a unified and practical pathway toward intelligent, high-efficiency 1-bit radar sensing and classification.

%%%%%%%%%%%%%%%%%%%%%%%%%%%%%%%%%%%%%%%%%%%%%%%%%%%%%%%%%%%%%%%%%%%%%%%%%%%%%%%%

%%%%%%%%%%%%%%%%%%%%%%%%%%%%%%%%%%%%%%%%%%%%%%%%%%%%%%%%%%%%%%%%%%%%%%%%%%%%%%%%
%                                REFERENCES
%%%%%%%%%%%%%%%%%%%%%%%%%%%%%%%%%%%%%%%%%%%%%%%%%%%%%%%%%%%%%%%%%%%%%%%%%%%%%%%%
% References section
\bibliographystyle{IEEEtran}
\bibliography{myreferences} % <-- This tells LaTeX to use your 'myreferences.bib' file

\ifCLASSOPTIONcaptionsoff
  \newpage
\fi

\end{document}